\documentclass[pre,twocolumn,amsmath,showpacs,superscriptaddress,nofootinbib]{revtex4}

\usepackage{rotating,longtable,color}
\usepackage{url}
\usepackage[colorlinks,linkcolor=blue,citecolor=magenta,urlcolor=blue]{hyperref}
\usepackage{txfonts}

\def\lb{\left}
\def\rb{\right}
\def\nn{\nonumber}
\def\ER{Erd\H{o}s-R\'{e}nyi}
\def\Pkk{$P(k,k')$}
\def\pk{p_k}
\def\Dbar{\bar D}
\def\DD{distribution of distances}
\def\qbar{\bar q}

\begin{document}

\title{Simple and accurate analytical calculation of shortest path lengths}

\author{Sergey Melnik}
\affiliation{MACSI, Department of Mathematics \& Statistics, University of Limerick, Ireland}
\author{James P. Gleeson}
\affiliation{MACSI, Department of Mathematics \& Statistics, University of Limerick, Ireland}


\begin{abstract}
We present an analytical approach to calculating the distribution of shortest paths lengths (also called intervertex distances, or geodesic paths) between nodes in unweighted undirected networks. We obtain very accurate results for synthetic random networks with specified degree distribution (the so-called configuration model networks). Our method allows us to accurately predict the distribution of shortest path lengths on real-world networks using their degree distribution, or joint degree-degree distribution. Compared to some other methods, our approach is simpler and yields more accurate results. In order to obtain the analytical results, we use the analogy between an infection reaching a node in $n$ discrete time steps (i.e., as in the susceptible-infected epidemic model) and that node being at a distance $n$ from the source of the infection.
\end{abstract}
\maketitle

\section{Introduction} \label{sec1}
The shortest path length between two nodes is the number of edges in the shortest path between them. The distribution of shortest path lengths and the average shortest path length of the network are important measures of the network topology~\cite{Newman10,Dorogovtsev08} as they characterize the efficiency of various spreading processes on networks; the analysis of shortest path lengths is at the centre of the six degrees of separation and the small-world phenomena~\cite{Watts98}. The calculation of shortest path lengths have been also used for estimating the accuracy of analytical approximations for dynamics on networks~\cite{Melnik11}, examining the onset of synchronization~\cite{zhao06} and assessing the resilience of communication networks to attacks and failures~\cite{Albert00}.

Significant effort has been devoted to the development of efficient numerical algorithms for both the exact and approximate calculation of the intervertex distances on a given network (see, for example,~\cite{Zwick01} and related literature). Exact numerical calculation of the probability distribution of distances between a pair of randomly chosen nodes (which requires to solve all-pairs shortest path problem) using the well-known Dijkstra algorithm has running time $O(mN + N^2\log N)$, where $N$ is the number of nodes and $m$ is the number of edges of the graph. Although there has been continuous improvement, the numerical calculation of shortest path lengths in large networks will remain a computationally expensive task.

Relatively little attention has been paid to the analytical calculation of distances on random networks~\cite{Newman01a,Dorogovtsev03b, Fronczak04,Fronczak04b,Dorogovtsev06b,Dorogovtsev08b,Katzav15}. In this paper, we consider undirected unweighted random networks with prescribed degree distribution $\pk$ or joint degree-degree distribution \Pkk~and propose a simple analytical method for calculating the probability distribution of shortest path lengths. Our method is more accurate than some other analytical methods for such classes of networks and allows us to predict (very often more accurately than any other know analytical method) shortest path lengths in real-world networks from their degree distribution $\pk$ or joint degree-degree distribution \Pkk. 

This paper is organized as follows. In Sec.~\ref{s:analogy}, we formulate the calculation of shortest path lengths in terms of a susceptible-infected epidemic model. In Sec.~\ref{s:z_regular}, we explain our analytical approach using $z$-regular random networks, and generalize it to random networks with arbitrary degree distribution or with degree-degree correlations in Secs.~\ref{pk_theory} and \ref{Pkk_theory} respectively. We apply our approach to calculate intervertex distances in real-world networks in Sec.~\ref{s:RWNs}, and conclude in Sec.~\ref{s:conclusions}.

\section{Analogy between intervertex distance and time to infection} \label{s:analogy}
\begin{figure}[b]
\centering
\includegraphics[width=0.7\columnwidth]{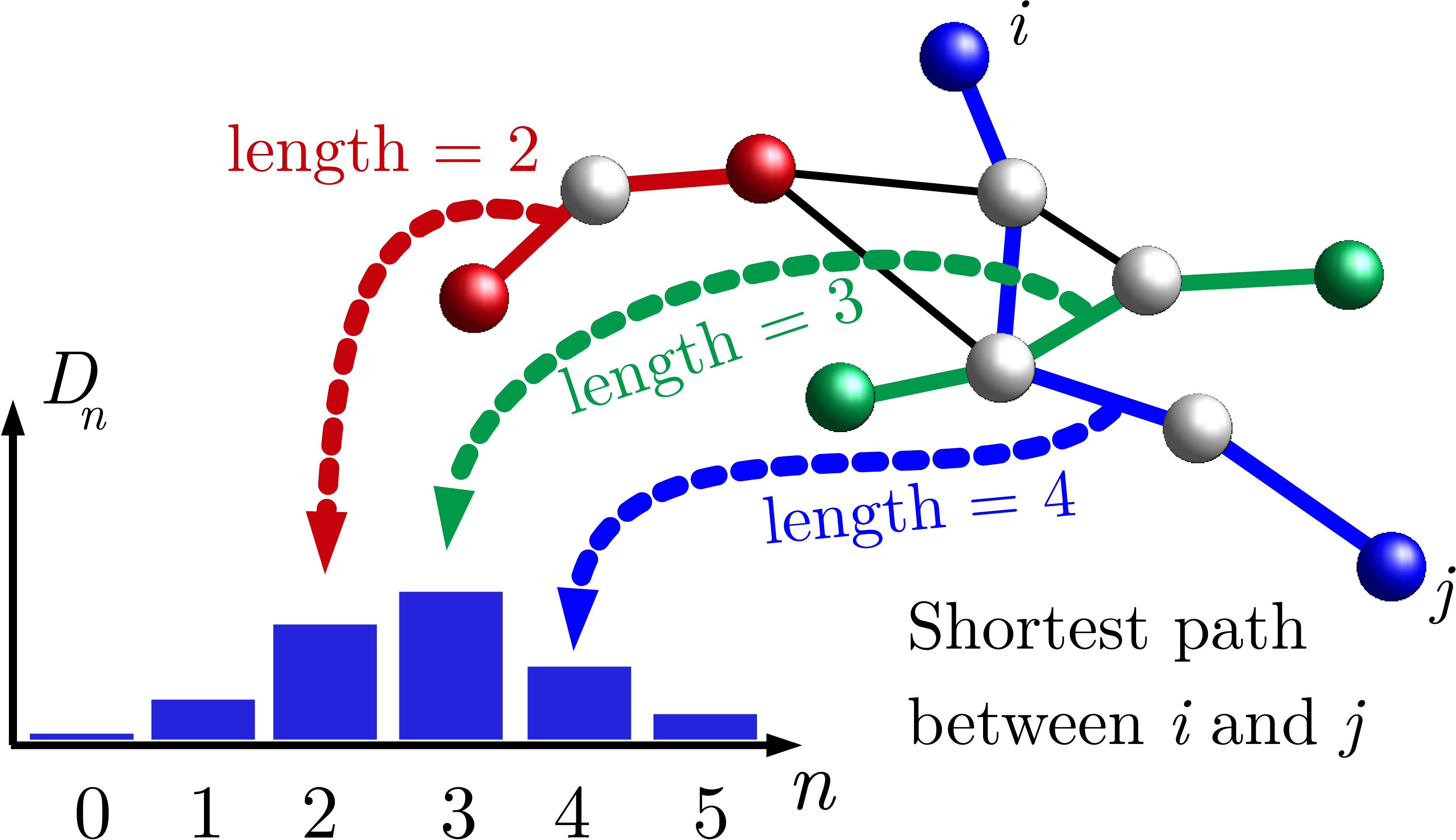}
\caption{(Color online) For a given network, the probability $D_n$ that two randomly chosen nodes are at a distance $n$ can be obtained numerically by calculating the shortest path length between each pair of nodes. For example, the shortest path between nodes $i$ and $j$ has length $4$ (blue links), so it contributes to $D_4$.}
\label{f:Dn_APSP_Hist}
\end{figure}

Let $D_n$ be the probability that the length of the shortest path (the distance) between two randomly-chosen nodes is equal to $n$. To calculate it numerically for a given network one could go through all pairs of nodes, find the shortest path between each pair and build a histogram as shown in Fig.~\ref{f:Dn_APSP_Hist}. 

\begin{figure}[h]
\centering
\includegraphics[width=0.7\columnwidth]{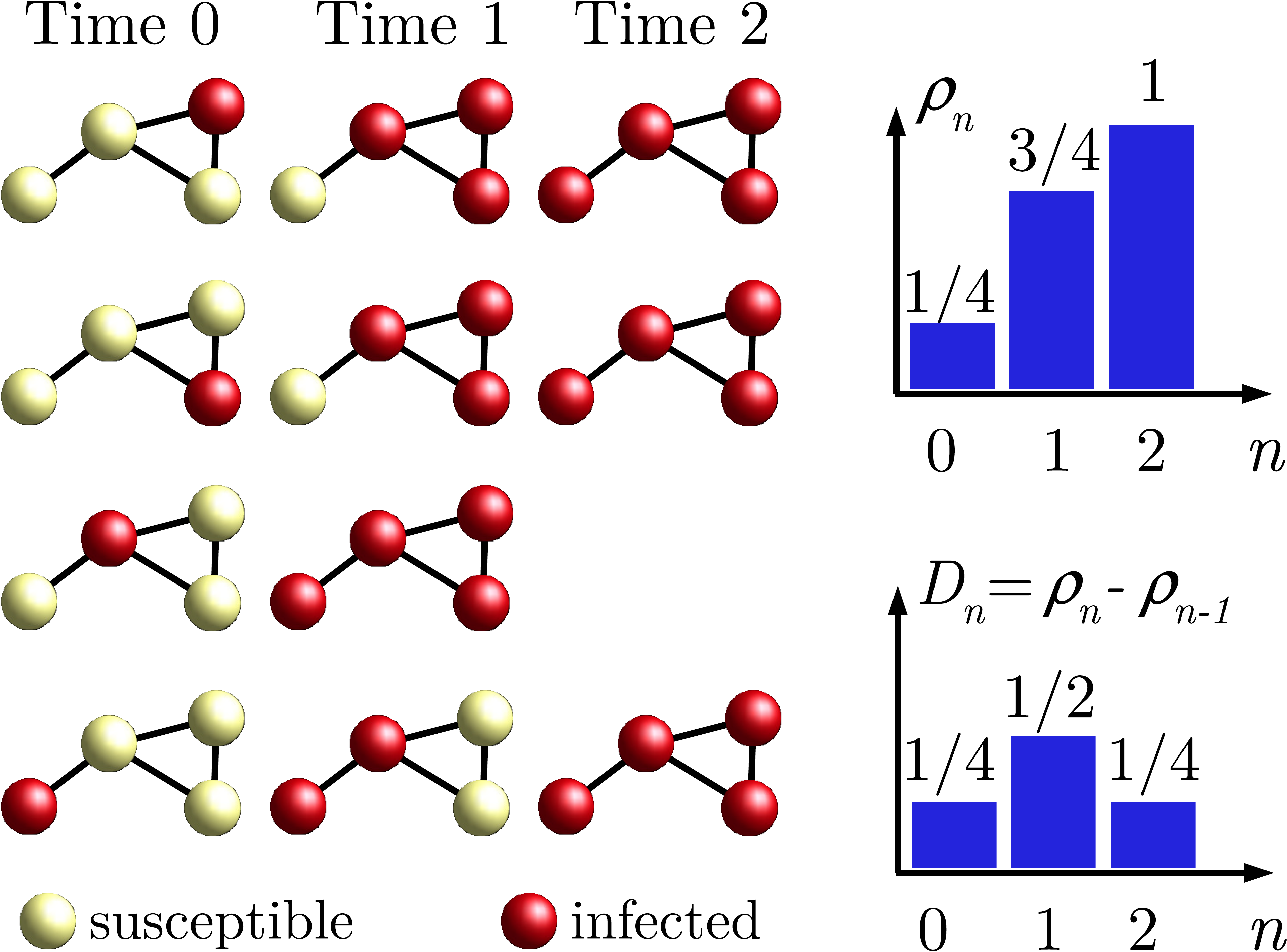}
\caption{(Color online) (Left) A single node (the seed) is infected at Time 0, and at each subsequent time step, every infected node infects all its susceptible neighbors. In this process, the time when other nodes become infected gives the distance between those nodes and the seed. (Top right) Let $\rho_n$ be the expected fraction of infected nodes at time $n$ as we run this epidemic process for different seeds. Here, $\rho_1 = (\frac34+\frac34+\frac44+\frac24)/4 = 3/4$. (Bottom right) Importantly, $\rho_n$ determines the \DD~between two randomly-chosen nodes: $D_n= \rho_n-\rho_{n-1}$.}
\label{f:SI_epidemic}
\end{figure}

To proceed with our analytical approach, let us first formulate the calculation of \DD~in terms of the following susceptible-infected epidemic model. Let us start with a single infected node $i$ as the source of an epidemic. At each discrete time step $n$, let every infected node infect all of its susceptible neighbors. Infected nodes remain infected indefinitely.\footnote{One can think of this process as susceptible-infected epidemic with infection probability 1, or as the Watts threshold model~\cite{Watts02} with threshold $\phi=0$. In both cases the states of all nodes are updated at each discrete time step.} At step $n=1$, node $i$ infects all its immediate neighbors, at step $n=2$ the second neighbors of $i$ become infected, and so on. The process stops when all nodes in the network become infected, see Fig.~\ref{f:SI_epidemic}. Notice that the distribution of distances from node $i$ to other nodes is given by the distribution of times when other nodes became infected. For example, the nodes who became infected at time $2$ are exactly $2$ steps away from node $i$. Let $\rho^i_n$ be the fraction of network nodes infected at step $n$ when node $i$ is initially infected. 

Repeat this epidemic process for every node $i$ in the network (i.e., each time starting from a different initially infected node $i$) as shown in Fig.~\ref{f:SI_epidemic}, and let $\rho_n \equiv \lb<\rho^i_n\rb>$ be the expected fraction of infected nodes at time $n$. Then the distribution of intervertex distances is given by the difference between $\rho_n$ and $\rho_{n-1}$
\begin{align}
\label{Dn}
D_n = \rho_n - \rho_{n-1},
\end{align}
and the average shortest path length is $\Dbar=\sum_n n D_n$.

Therefore, if we analytically calculate the expected fraction of infected nodes at time $n$ for the above epidemic process, we can obtain the distribution of shortest path lengths. Below we explain the main idea of our approach using, for simplicity, regular random graphs (i.e., networks where every node has $z$ neighbors and connections between nodes are random), and in the subsequent sections we generalize the results to random networks with arbitrary degree distribution (the so-called configuration model networks) and to networks with degree-degree correlations.

\section{Explaining our approach using $z$-regular random graphs}\label{s:z_regular}
We consider a randomly chosen node $A$ and calculate its probability of being infected at time $n$ (i.e., after $n$ synchronous updates of states of all nodes). This probability is $\rho_n$ since we chose $A$ uniformly at random. Initially a single node $i$ in the network is infected, which implies that $A$ is initially infected with probability $\rho_0=1/N$, where $N$ is the total number of nodes, or initially susceptible with probability $1-\rho_0$.

We denote by $q_{n}$ the probability that at time $n$ (i.e., immediately before update $n+1$ of node $A$) a random neighbor $B$ of node $A$ is infected, conditioned on node $A$ itself being susceptible. This conditioning accounts for the fact that neighbors of $A$ become infected before $A$, and that $A$ did not infect them. Note that $A$ is susceptible only if it was initially susceptible and none of its neighbors have yet infected it. Thus, the probability that $A$ is infected at time $n+1$ is
\begin{align}
\label{RRG_rhon}
\rho_{n+1} = 1 -(1-\rho_0) (1-q_{n})^z,
\end{align}
where $(1-\rho_0)$ is the probability that $A$ is initially susceptible, and $(1-q_{n})^z$ is the probability that none of $z$ neighbors of $A$ is infected at time $n$.\footnote{In Eq.~\eqref{RRG_rhon}, we have assumed that the states of any two neighbors of node $A$ are independent, which is the case for a graph that is locally tree-like, such as random networks constructed using the configuration model~\cite{Newman10}. Although the tree-like assumption breaks down on real-world networks with high clustering coefficients and/or significant community structure, we have previously demonstrated that results obtained using locally tree-like approximations often remain reasonably accurate on such networks~\cite{Melnik11,Gleeson12}.}
\begin{figure}[h]
\centering
\includegraphics[width=0.8\columnwidth]{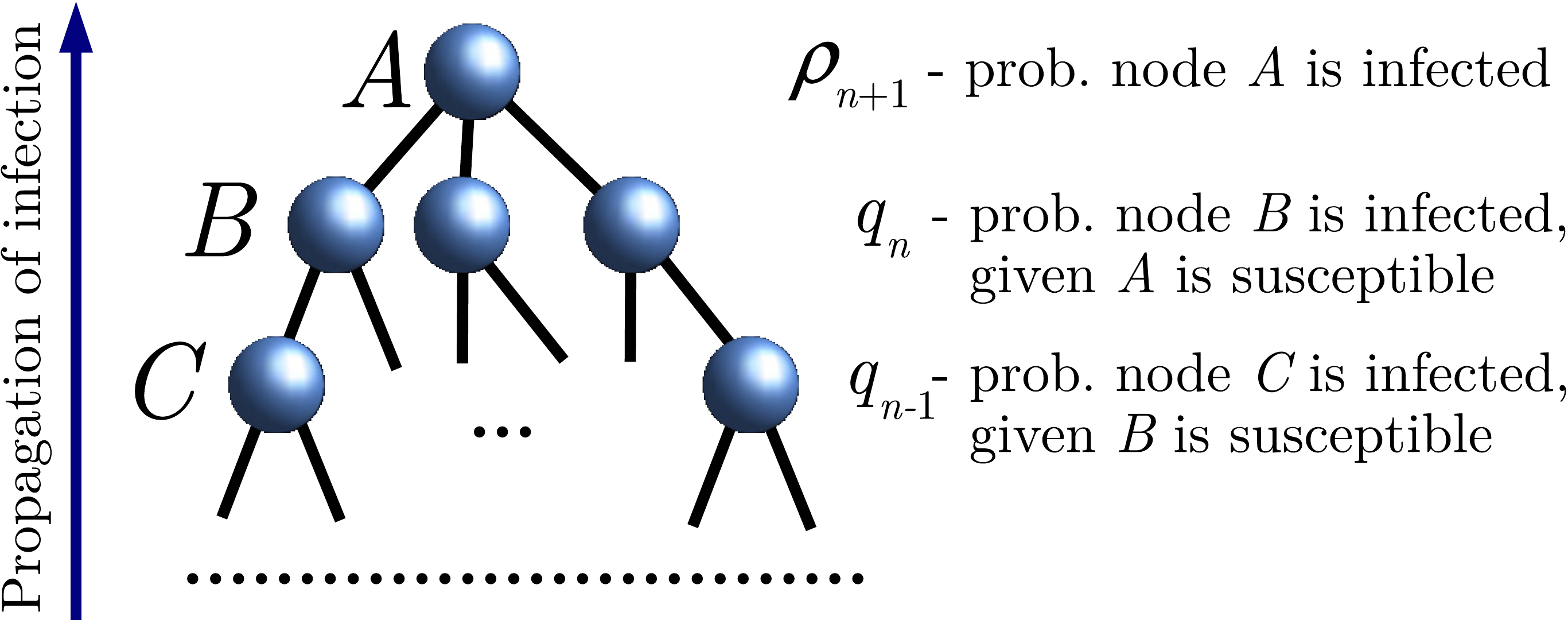}
\caption{(Color online) Tree-like structure of a 3-regular random network near node $A$. Since $A$ is a randomly chosen node, the probability that $A$ is infected at time $n+1$ is $\rho_{n+1}$. Here, $q_n$ is the probability that node $B$, a child of $A$, is infected by any of its children, given $A$ is susceptible. Similarly, $q_{n-1}$ is the probability that $C$ (a child of susceptible $B$) is infected by any of $C$'s own children.}
\label{f:tree_update}
\end{figure}

In order to calculate $q_n$, we consider node $B$, a neighbor of $A$, and establish a recurrence relation for $q_n$~\cite{Melnik13}. Using similar reasoning as for Eq.~\eqref{RRG_rhon}, we express the probability $q_n$ (that $B$ is infected, given $A$ is susceptible) in terms of probability $q_{n-1}$ that a child of $B$, node $C$, --- defined as a neighbor of $B$ that is one step further away from $A$ --- is infected given $B$ is susceptible, see Fig.~\ref{f:tree_update}:
\begin{align} 
\label{RRG_qn}
q_{n} = 1-(1-\rho_0)(1-q_{n-1})^{z-1}.
\end{align}
The power $z-1$ in this equation appears instead of $z$ as in Eq.~\eqref{RRG_rhon} because, by the definition of $q_n$, node $A$ is susceptible and cannot infect $B$ and thus is excluded from consideration (likewise $B$ cannot infect its children, and so on).

\begin{figure}[h]
\centering
\includegraphics[width=0.97\columnwidth]{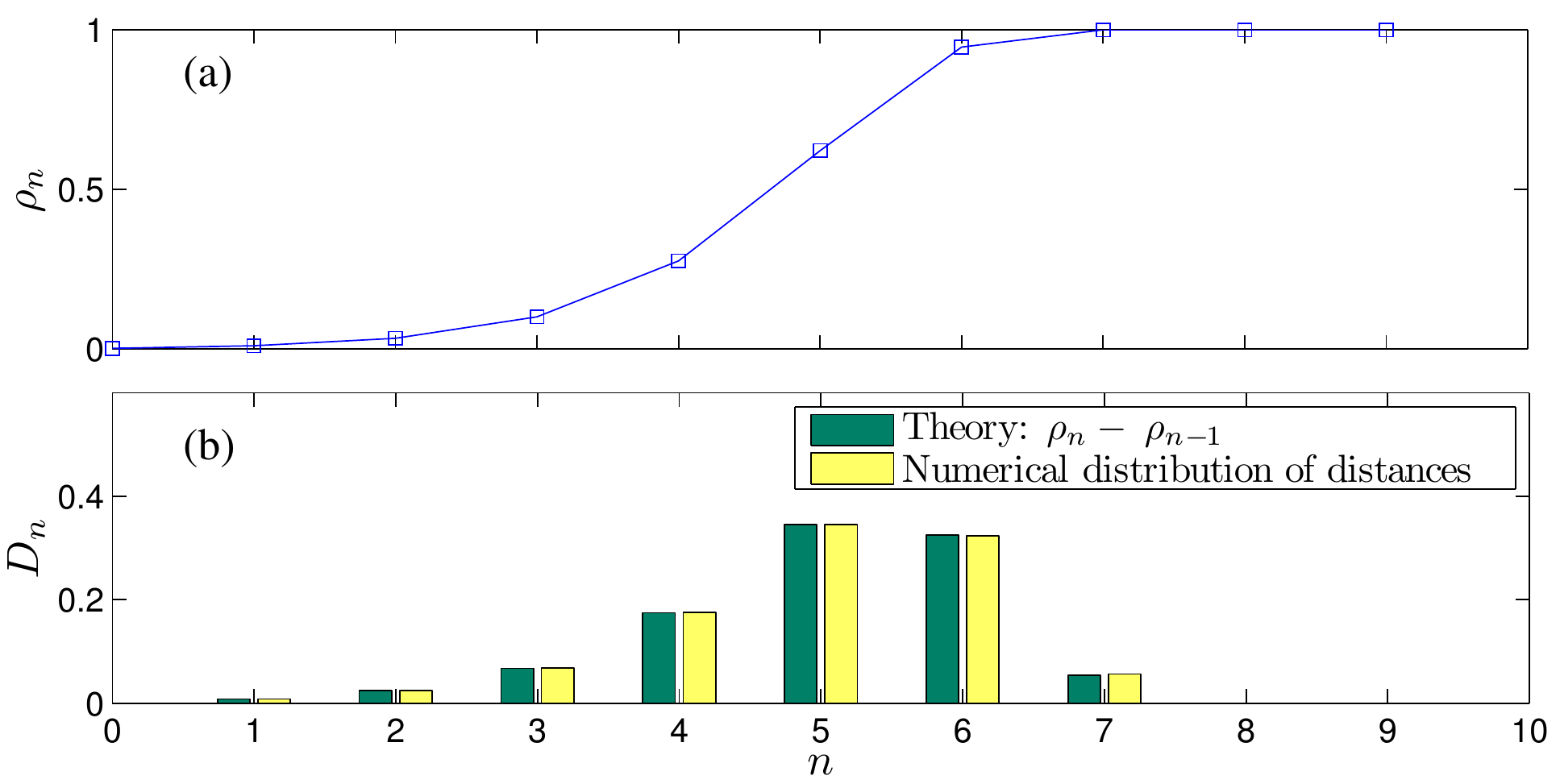}
\caption{(Color online) (a) Time evolution of the fraction of infected nodes $\rho_n$ for the discrete-time susceptible-infected epidemic model described in the text, calculated by iterating Eqs.~\eqref{RRG_rhon}-\eqref{RRG_qn} starting with $q_0 = \rho_0 = 1/N$, for 4-regular random network with $N=500$ nodes. (b) Values of $\rho_n-\rho_{n-1}$ together with numerically calculated \DD. Numerical results are averaged over 10 realizations of networks.}
\label{f:rho_n_Dn_vs_n}
\end{figure}

Therefore, starting with $q_0 = \rho_0=1/N$, we can iterate Eqs.~\eqref{RRG_rhon}-\eqref{RRG_qn} to obtain values of $\rho_n$, and use them in Eq.~\eqref{Dn} to calculate the \DD. In Fig.~\ref{f:rho_n_Dn_vs_n}, we show the values of $\rho_n$ calculated for a 4-regular random network with $N=500$ nodes, and the corresponding values of $D_n$. Observe the excellent match between numerical and theoretical results.

\begin{figure}[h]
\centering
\includegraphics[width=0.97\columnwidth]{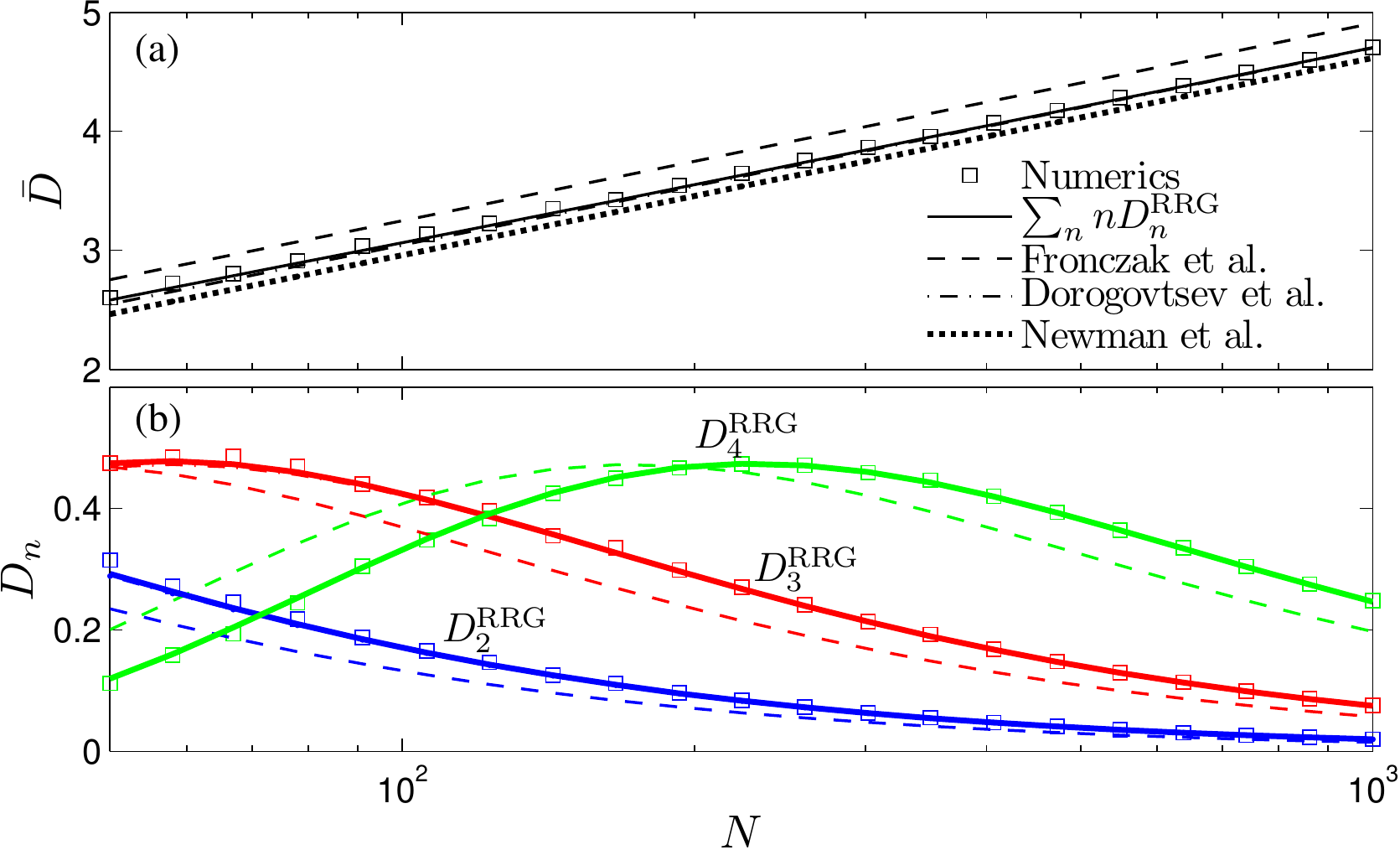}
\caption{(Color online) (a) Average shortest path length $\Dbar=\sum_n n D_n$ and (b) the probabilities $D_n$ that the distance between a random pair of nodes is $n$ (for $n\in\{2,3,4\}$) as functions of the number of network nodes $N$ for a regular random graph with mean degree $z=5$. We compare the results of our Eq.~\eqref{LargeNRRG} and the analytical results of~\cite{Fronczak05,Dorogovtsev03b,Newman01a} (presented in Appendices~\ref{App:RRG} and \ref{App:Davg}) with direct numerical simulations shown by symbols. In all cases Eq.~\eqref{LargeNRRG} provides the best prediction of numerical results, which remains accurate even for small $N$. The results of Dorogovtsev et al.~\cite{Dorogovtsev03b} match very closely to~Eq.~\eqref{LargeNRRG}, but are more difficult to calculate. Numerical results are averaged over 10 realizations of networks.}
\label{RRG_z5}
\end{figure}

We can analytically solve Eqs.~\eqref{RRG_rhon}-\eqref{RRG_qn} (see Appendix~\ref{App:RRG}) and obtain an explicit formula for $D_n$ for $z$-regular networks:
\begin{align}
D_n^{\rm RRG} = \exp \lb[ -\frac{z(z-1)^{n-1}-2}{(z-2)N} \rb] - \exp \lb[ -\frac{z(z-1)^n-2}{(z-2)N} \rb].
\label{LargeNRRG}
\end{align}
We show in Fig.~\ref{RRG_z5} that this expression predicts the numerical results extremely accurately even for $N$ as low as $50$, and outperforms some previous analytical results~\cite{Fronczak05,Dorogovtsev03b,Newman01a} which we present in Appendices~\ref{App:RRG} and \ref{App:Davg}. The accuracy observed in Fig.~\ref{RRG_z5} for regular random networks suggests potential high accuracy of our approach for more complicated network topologies that we consider next.

\section{$p_k$-theory: Generalization to random networks with arbitrary degree distribution} \label{pk_theory}
To generalize Eqs.~\eqref{RRG_rhon}-\eqref{RRG_qn} from $z$-regular random networks to networks with an arbitrary degree distribution $p_k$ constructed using the configuration model~\cite{Newman10}, we consider an epidemic started by a degree-$k'$ seed node, and let $\rho_n^{k,k'}$ be the corresponding expected fraction of degree-$k$ nodes infected at time step $n$. We can calculate $\rho_n^{k,k'}$ using the following set of recurrence equations
\begin{align}
\rho_{n+1}^{k,k'} &= 1-(1-\rho_0^{k,k'}) (1-\qbar_n^{k,k'})^k, \label{pk_rhon} \\
q_{n+1}^{k,k'} &= 1- (1-\rho_0^{k,k'})(1-\qbar_n^{k,k'})^{k-1}, \label{pk_qn} \\
\qbar_n^{k,k'} &= \sum_k \frac{k p_k}{z} q_n^{k,k'}, \label{pk_qbar}
\end{align}
with initial values 
\begin{align}\label{rho0}
 q_0^{k,k'} = \rho_0^{k,k'}=\frac{\delta_{k,k'}}{N p_{k'}}.
\end{align}
Here $q_n^{k,k'}$ is the probability that a degree-$k$ node is infected, given that its parent is susceptible, and $\qbar_n^{k,k'}$ is the probability that a child of a susceptible degree-$k$ node is infected at time step $n$ of an epidemic started from a single infected node of degree $k'$,\footnote{Note that $\qbar_n^{k,k'}$ given by~\eqref{pk_qbar} is independent of $k$; we write it in this general form so that Eqs.~\eqref{pk_rhon}-\eqref{pk_qn} are compatible with Eq.~\eqref{qbar} of Sec.~\ref{Pkk_theory}.} and $z=\sum_k k p_k$ is the mean degree. Note that Eqs.~\eqref{pk_rhon}-\eqref{pk_qbar} reduce to Eqs.~\eqref{RRG_rhon}-\eqref{RRG_qn} for the $z$-regular case of $p_k=\delta_{k,z}$.

Since finite paths only exist in a connected component, we focus our further analysis on the giant connected component (GCC) of the network, and exclude small components and nodes with degree 0. Note that for configuration model networks, GCC is the entire network (in the thermodynamic limit) if and only if the degree distribution $p_k$ does not contain degree-0 and degree-1 nodes. If there are degree-1 nodes in the degree distribution, the generated network will have the fractional GCC size $<1$.\footnote{If $p_1>0$, pairs of connected degree-1 nodes, as well as other nodes or small components surrounded by degree-1 nodes will exist and will not belong to GCC.} For example, for \ER~networks with low mean degree $z$, GCC size is significantly below 1, and asymptotically approaches $1$ as $z$ increases.

The values of $\rho_n^{k,k'}$ calculated by iterating Eqs.~\eqref{pk_rhon}-\eqref{pk_qbar} tell us the fractions of infected nodes in the entire network, as opposed to that in the GCC. Since all nodes in GCC eventually become infected, the steady state values $\rho_\infty^{k}\equiv\rho_\infty^{k,k'}\le 1$ (that do not depend on the seed node degree $k'$) give us the fraction of degree-$k$ nodes who are part of GCC~\cite{Melnik14,Gleeson08a}. Hence, the fraction of infected degree-$k$ nodes in GCC at time $n$ is $\rho_n^{k,k'} /\rho_\infty^{k,k'}$.
The probability that two nodes (chosen from the entire network) are not connected is $D_\infty = 1-(\sum_k p_k \rho_\infty^{k})^2$, where $\sum_k p_k \rho_\infty^{k}$ is the fractional size of GCC.
Next, assuming that a pair of nodes is chosen from GCC, the probability that two random nodes with degrees $k$ and $k'$ are at a distance $n$ from each other is
\begin{align}
 D_n^{k,k'} = \frac{\rho_n^{k,k'} - \rho_{n-1}^{k,k'}}{\rho_\infty^{k,k'}} \label{e:Dnkk};
\end{align}
the probability that a randomly chosen node (of any degree) is at a distance $n$ from a randomly chosen degree-$k$ node is
\begin{align}
 D_n^k=\sum_{k'} p_{k'} D_n^{k,k'} \label{e:Dnk},
\end{align}
and 
\begin{align}
 D_n=\sum_{k} p_{k} D_n^{k} \label{e:Dn}
\end{align}
is the probability that the distance between a pair of random nodes in GCC is $n$.

\begin{figure}[htb]
\flushright
\includegraphics[width=0.95\columnwidth]{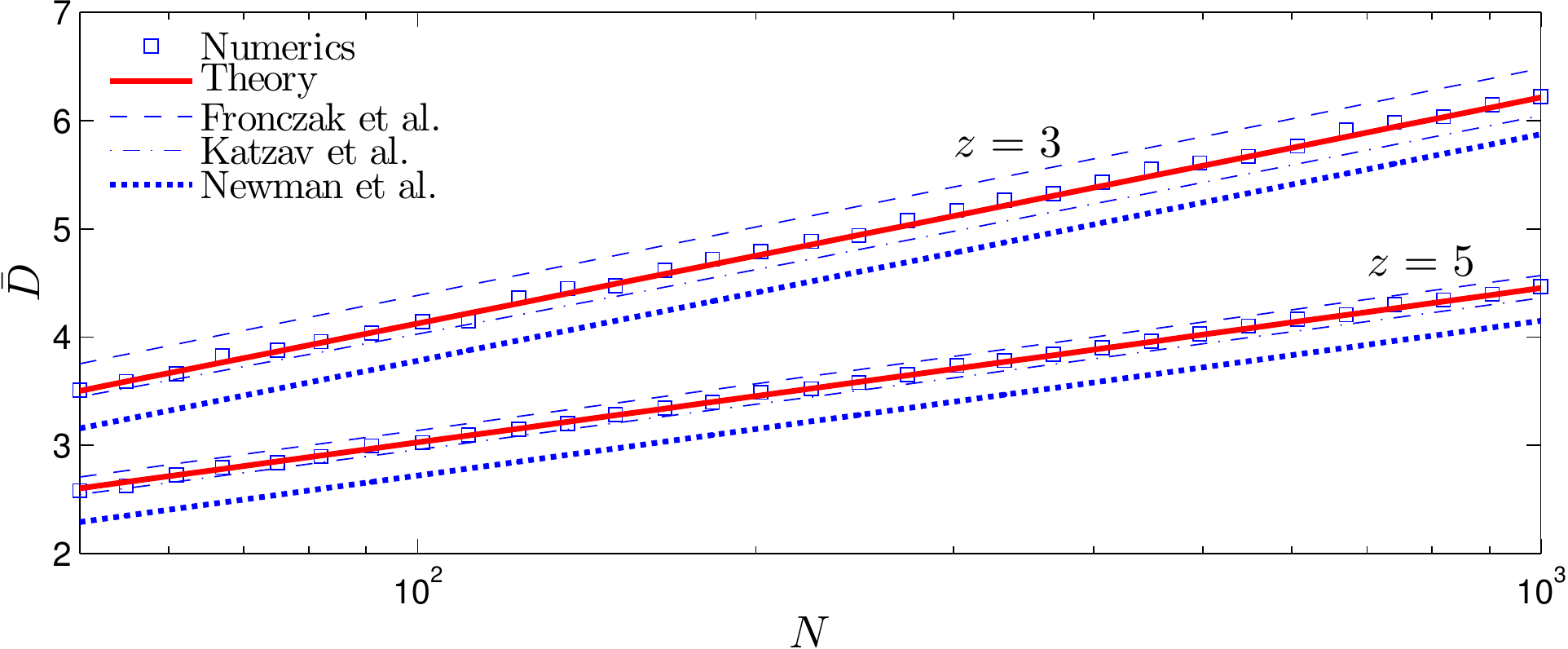}
\includegraphics[width=0.97\columnwidth]{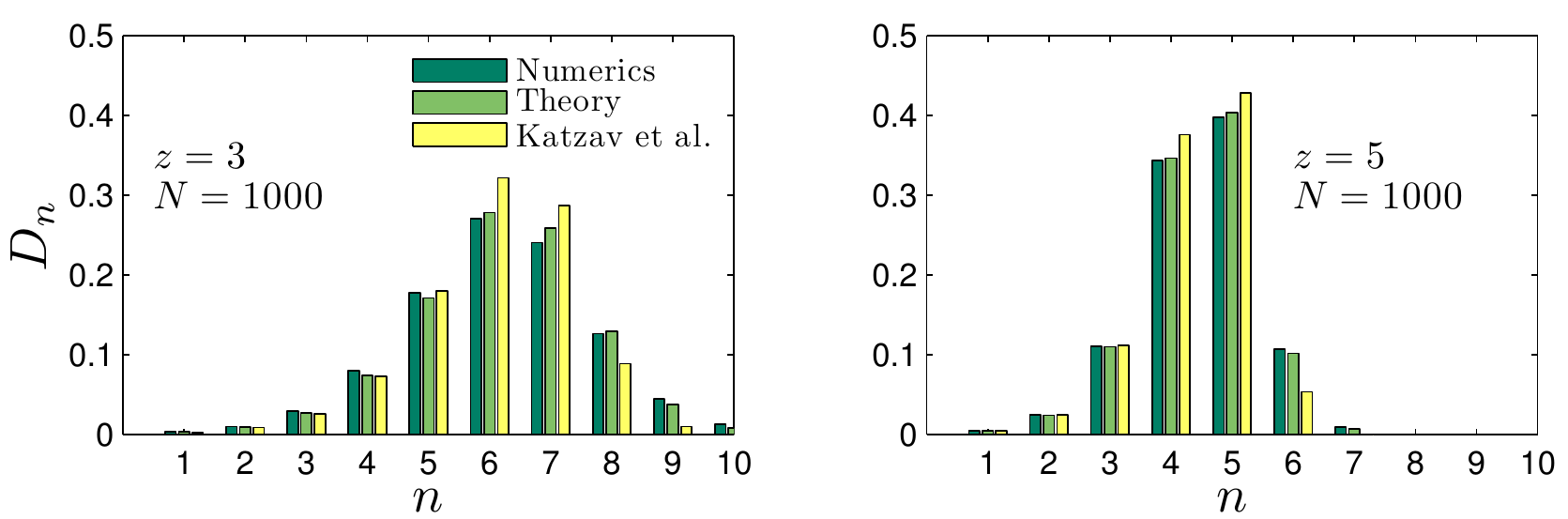}
\caption{(Color online) Average shortest path length $\Dbar=\sum_n n D_n$ versus the number of nodes $N$ (top panel) and the \DD~$D_n$ (bottom panels) for GCC of \ER~networks with mean degree $z=3$ and $z=5$. We compare our theoretical results (Eqs.~\eqref{pk_rhon}-\eqref{e:Dn}) and the analytical results of~\cite{Fronczak05,Newman01a,Katzav15} (presented in Appendices~\ref{App:Davg} and \ref{App:Katzav}) with direct numerical simulations. In all cases our theory provides the best prediction of numerical results, which remains accurate even for small $N$. Numerical results are averaged over 50 realizations of networks.}
\label{ERG}
\end{figure}
In Fig.~\ref{ERG}, we illustrate our approach using \ER~networks and compare the results with numerical simulations, and with some previously obtained analytical results~\cite{Fronczak05,Newman01a,Katzav15} presented in Appendices~\ref{App:Davg} and \ref{App:Katzav}. We use the Poisson degree distribution $p_k = e^{-z}z^k/k!$ of~\ER~networks in Eqs.~\eqref{pk_rhon}-\eqref{e:Dn} and plot the results in Fig.~\ref{ERG}. Our theory agrees better with numerical simulations than the previous analytical approaches. The prediction of the \DD~is excellent, but not as perfect as in the $z$-regular case in Fig.~\ref{f:rho_n_Dn_vs_n}(b); we explain the reasons for this in Appendix~\ref{s:error_pk}.

\section{$P(k,k')$-theory: Generalization to random networks with degree-degree correlations} \label{Pkk_theory}
Our approach can be easily generalized to random networks specified by the joint degree-degree distribution $P(k,k')$, which is defined as the probability that a randomly chosen network edge connects a degree-$k$ node to a degree-$k'$ node~\cite{Gleeson08a,Melnik11,Melnik14}. The joint distribution $P(k,k')$ determines the degree distribution of the network
\begin{align}
 p_k = \frac{\sum_{k'}P(k,k')/k}{\sum_{k',k''}P(k',k'')/k'},
\end{align}
but it also contains additional information about the correlation of node degrees at either end of an edge. Thus, $P(k,k')$ describes the network topology more accurately than $p_k$, and using $P(k,k')$ should improve the accuracy of our approach when we apply it to real-world networks. 

Equations~\eqref{pk_rhon}-\eqref{e:Dn} of Sec.~\ref{pk_theory} can be directly applied to networks specified by the joint degree-degree distribution $P(k,k')$, except that Eq.~\eqref{pk_qbar} should be replaced with
\begin{align}
\qbar_n^{k,k'} = \frac{\sum_{k''}P(k,k'')q_n^{k'',k'}}{\sum_{k''}P(k,k'')} \label{qbar}.
\end{align}
In the case of degree-uncorrelated networks, the joint degree-degree distribution factorizes as $P(k,k') = k p_k k' p_{k'}/ z^2$ (since the degrees of nodes at either end of an edge are independent) and Eq.~\eqref{qbar} reduces to Eq.~\eqref{pk_qbar}.

\section{Application to Real-World Networks}\label{s:RWNs}
To apply our approach to real-world networks, we calculate the degree distribution $p_k$ and/or the joint degree-degree distribution $P(k,k')$ from the network adjacency matrix. We then use one or both of these distributions in the equations presented in Secs.~\ref{pk_theory} and \ref{Pkk_theory} to obtain theoretical results. We will refer to the results of our approach where we use $p_k$ of the network and Eq.~\eqref{pk_qbar} as $p_k$-theory, and where we use $P(k,k')$ of the network and Eq.~\eqref{qbar} as $P(k.k')$-theory.

\begin{figure}[h]
\centering
\includegraphics[width=0.97\columnwidth]{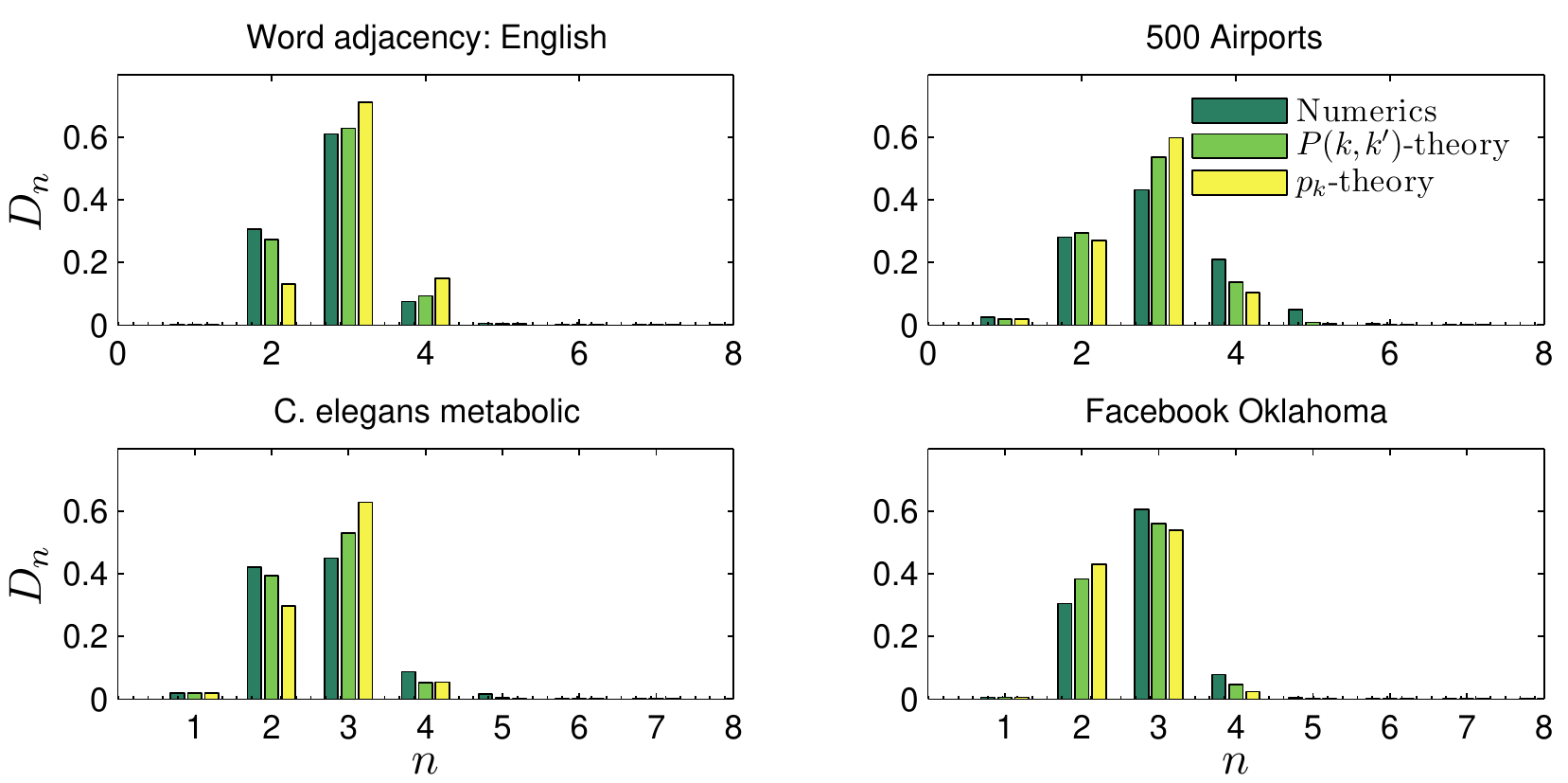}
\caption{(Color online) Distribution of distances $D_n$ for several real-world networks. Generally, $P(k,k')$-theory predicts the actual $D_n$ better than $p_k$-theory.}
\label{RWN}
\end{figure}

In Fig.~\ref{RWN}, we compare the numerically calculated \DD~for several real-world networks (see Table~\ref{table1}) with the results of $p_k$ and \Pkk-theories. In general, \Pkk-theory predicts the numerical \DD~better than $p_k$-theory as it uses additional information about the degree correlations in the network.

\setlength{\tabcolsep}{0.05cm}
\begin{table}[floatfix]
\begin{center}
\begin{tabular}{l|l r c c |c c c c}
\hline
&&&&& \multicolumn{4}{c}{Relative errors in $\Dbar$ for}\\
&Network&$N$&$z$&$\Dbar$&$p_k$&\Pkk&FR.& NMN.\\
\hline
\begin{rotate}{90}
\hspace{-2cm}Real world
\end{rotate}
&Word adj. Eng.~\cite{Onnela12}& 7377 & 12        &  2.78  & -0.09 & -0.02 & -0.04 & 0.24\\
&Word adj. French~\cite{Onnela12}& 8308 & 5.7 & 3.22  & -0.02 & 0.00 & 0.06 & 0.27\\
&500 Airports~\cite{Colizza07}& 500   & 11.9        & 2.99  & 0.07 & 0.06 & 0.08 & 0.35\\
&Interact. Prot.~\cite{Colizza05,Colizza06}&4713&6.0	& 4.22  & 0.08 & 0.06 & 0.10  & 0.28\\
&C. Eleg. Met.~\cite{Duch05} & 453	& 8.9	& 2.66	&-0.02 & 0.02 & 0.01 & 0.23\\
&C. Eleg. Neur.~\cite{Watts98}& 297	& 14.5	& 2.46  & 0.00 & 0.03 & 0.00 & 0.22\\
&FB Caltech~\cite{Traud08} & 762   & 43.7	& 2.34  & 0.03 & 0.03 & 0.03 & 0.29\\
&FB Georgetown~\cite{Traud08}& 9388  & 90.7	& 2.76  & 0.10 & 0.08 & 0.09  & 0.31\\
&FB Oklahoma~\cite{Traud08}& 17420 &102.5	& 2.77  & 0.07 & 0.04 & 0.07 & 0.29\\
\hline
\begin{rotate}{90}
\hbox{\hspace{-1.3cm}Synthetic}
\end{rotate}
% Relative errors       \Dbar &  pk  & Pkk &  FR  &  NM-R
&$z$-regular &500&4    & 5.02 & 0.00 & 0.00 & 0.07 & 0.00\\
&$z$-regular &500&10   & 2.96 & 0.00 & 0.00 & 0.02 & 0.08\\
&\ER &500&10           & 2.96 & 0.00 & 0.00 & 0.01 & 0.10\\
&$z$-regular &10000&4  & 7.73 & 0.00 & 0.00 & 0.05 & 0.00\\
&$z$-regular &10000&10 & 4.34 & 0.00 & 0.00 & 0.01 & 0.06\\
&\ER &10000&10         & 4.26 & 0.00 & 0.00 & 0.01 & 0.07
\end{tabular}
\end{center}
\caption{Comparison of the actual value of the mean distance $\Dbar$ for several real-world and synthetic networks with the values calculated using different analytical methods. Here, $N$ is the number of network nodes and $z$ is the mean degree. We show the relative errors $(\Dbar - \Dbar_{\rm theor.} )/ \Dbar$ for the four theories: our $p_k$ and \Pkk-theories of Secs.~\ref{pk_theory} and~\ref{Pkk_theory} respectively, (FR.) Eq.~\eqref{e:Dbar_FR} of Fronczak et al.~\cite{Fronczak05}, and (NMN.) Eq.~\eqref{e:Dbar_NMN} of Newman et al.~\cite{Newman01a}.
}
\label{table1}
\end{table}

\begin{figure}[h]
\flushleft
\includegraphics[width=0.98\columnwidth]{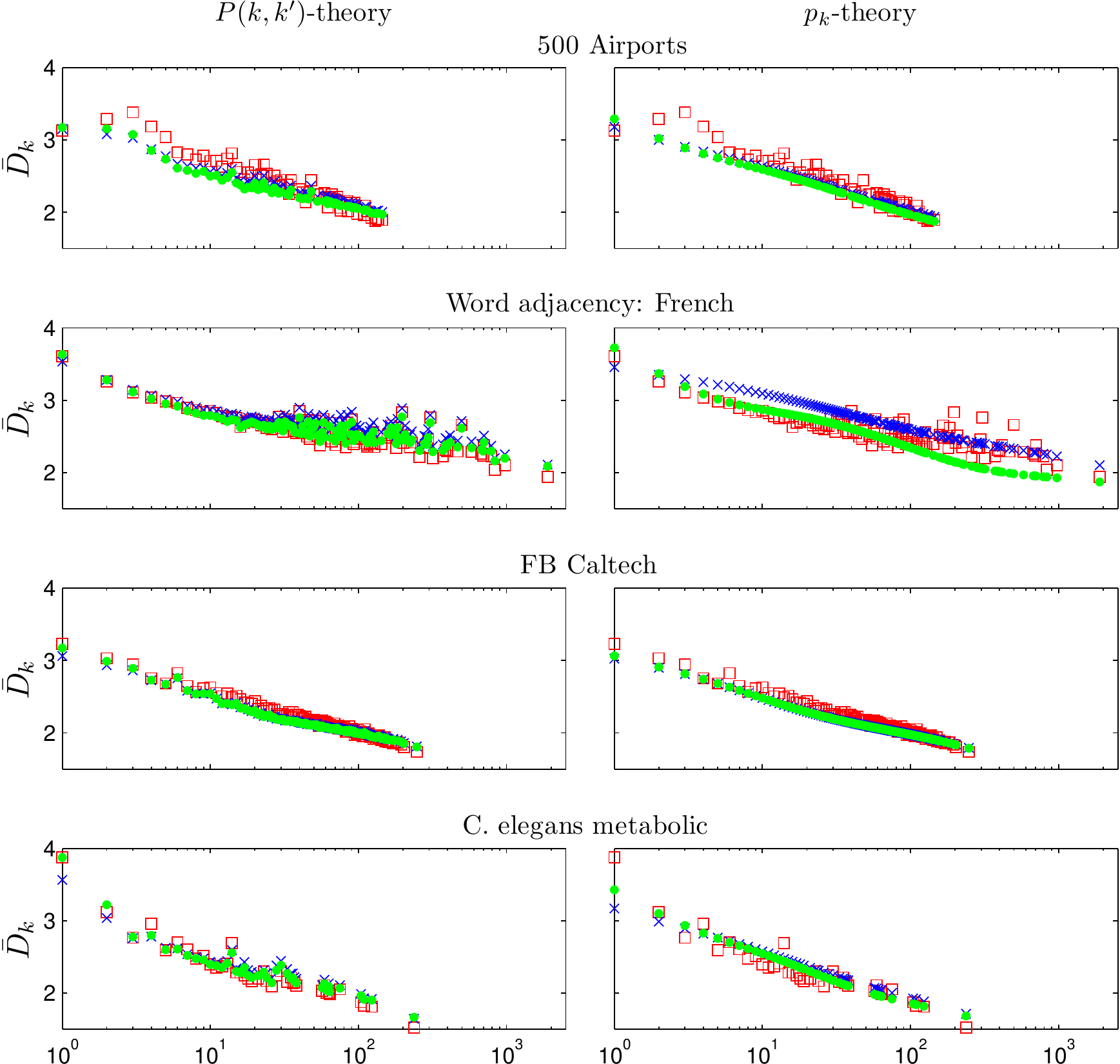}
\includegraphics[width=1\columnwidth]{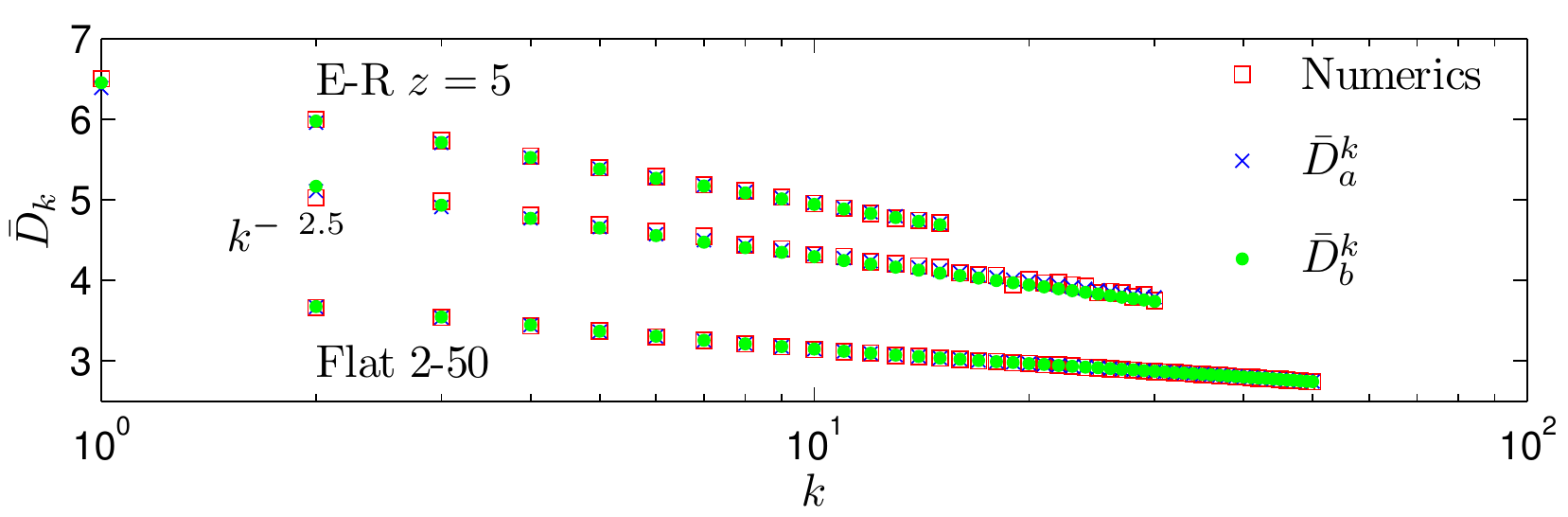}
\caption{(Color online) The average distance from a degree-$k$ node to all other nodes as a function of $k$. For real-world networks, the result of \Pkk-theory (left column) is more accurate than the result of $p_k$-theory (right column). The bottom panel shows the result for synthetic networks with $N=5000$ nodes: an \ER~graph with mean degree $z=5$, a random network with a (truncated) power-law degree distribution $p_k\propto k^{-2.5}$ ($2\le k\le 30$), and a random network with a ``flat'' degree distribution $p_k = \rm{const}$ ($2\le k\le 50$).}
\label{fig:Davg_vs_k}
\end{figure}

In Fig.~\ref{fig:Davg_vs_k}, we plot the expected length of the shortest path between two nodes one of which has degree $k$. We plot $\Dbar^{k}_a=\sum_{k',n} n p_{k'} D_{n}^{k',k}$ and $\Dbar^k_b=\sum_{k',n} n p_{k'} D_{n}^{k,k'}$. Here, $\Dbar^{k}_a$ is calculated based on an epidemic started from a degree-$k$ seed and averaged over the degrees of other nodes, while in $\Dbar^{k}_b$ we consider a degree-$k$ node and average over multiple epidemics started with seeds of various degrees. Of course these are the same when one runs numerical simulations of the epidemic process on a given finite network. However, our theoretical approach gives only approximate equality between $D_n^{k,k'}$ and $D_n^{k',k}$ (due to the infinite network assumption as we discuss in Appendix \ref{s:asymmetry}), leading to small differences between $\Dbar^{k}_a$ and $\Dbar^{k}_b$. We show the results for several real-world and synthetic networks. For real-world networks, \Pkk-theory usually better predicts the exact numerical results than $p_k$-theory. The errors for $\Dbar^{k}_a$ and $\Dbar^{k}_b$ are similar, and the choice of quantity that best matches the numerical results depends on a particular network. The errors may be attributed in part to clustering~\cite{Faqeeh15a,Melnik11}, modular structure~\cite{Melnik14}, or other topological features~\cite{Faqeeh16} present in these networks. For synthetic networks the results of \Pkk~and $p_k$-theories coincide because of the absence of degree-degree correlations, and are in excellent agreement with numerical results.

\begin{figure}[h]
\flushleft
\includegraphics[width=0.97\columnwidth]{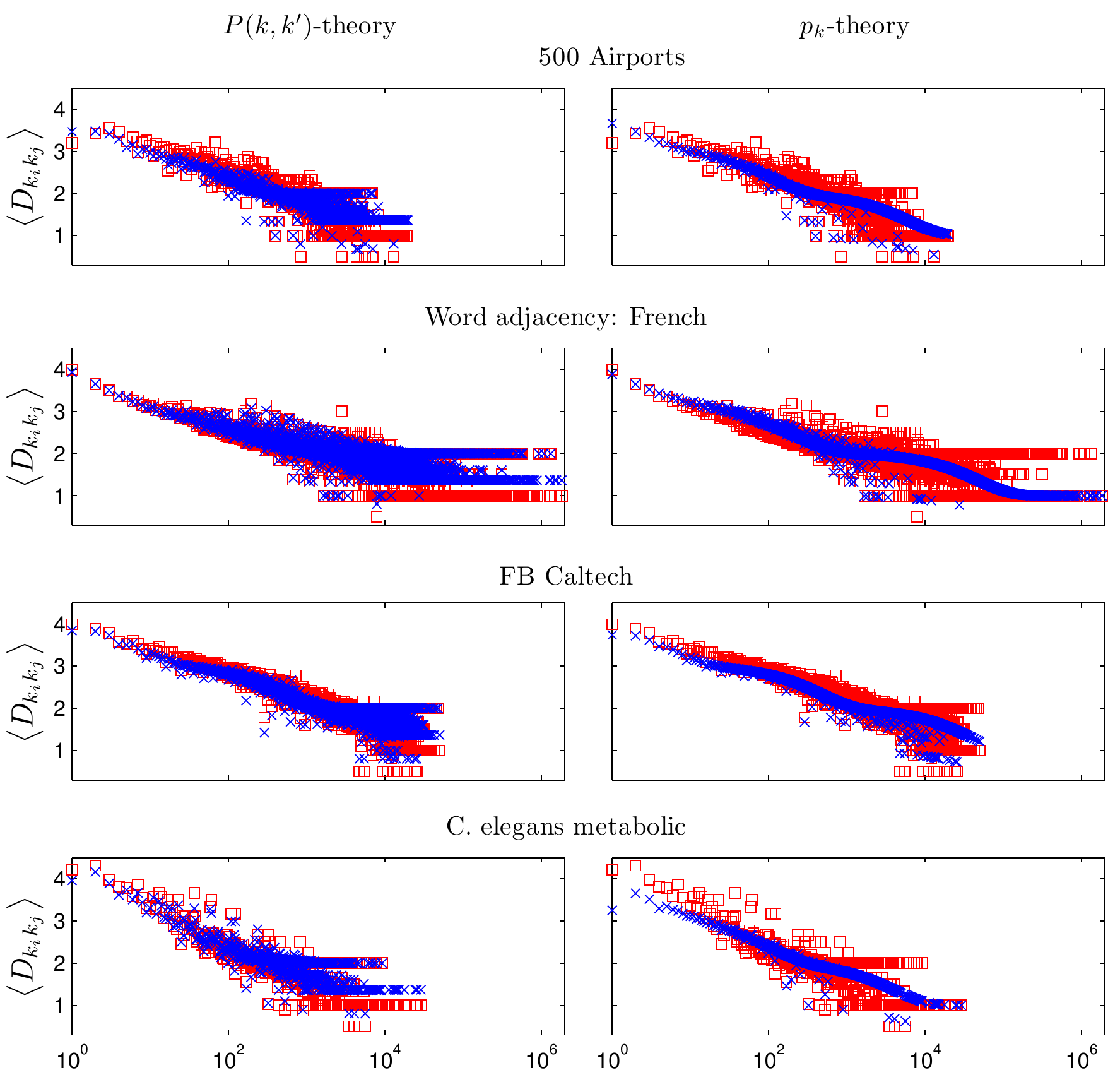}
\includegraphics[width=0.97\columnwidth]{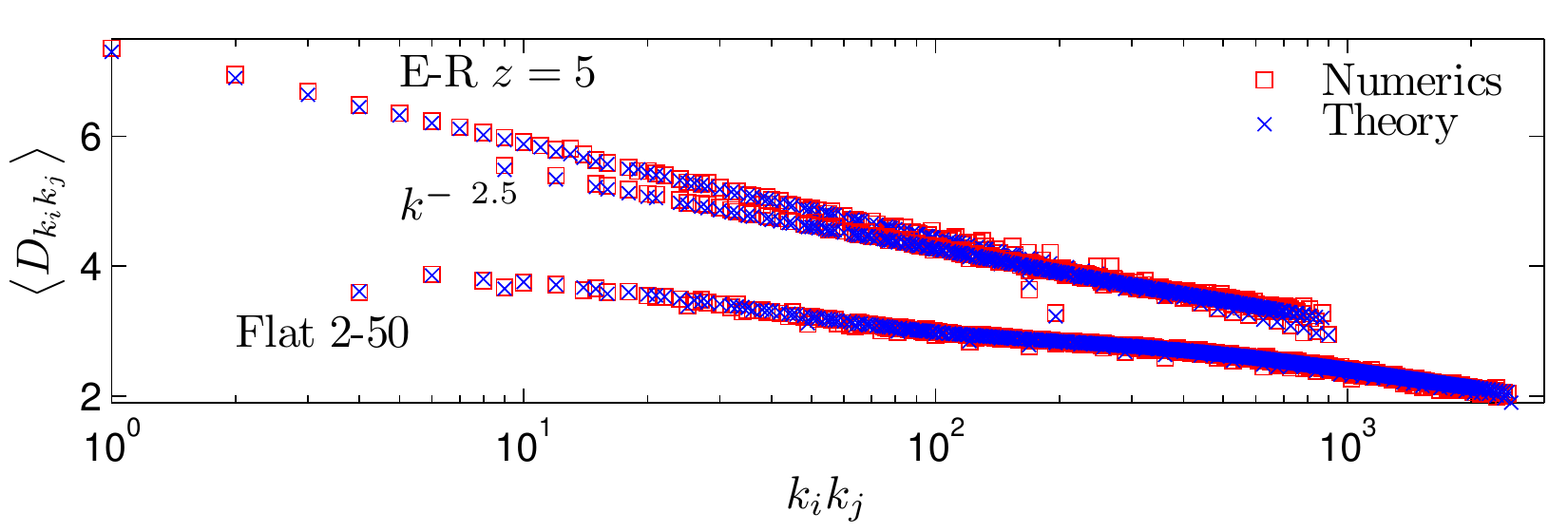}
\caption{(Color online) Average shortest path length between a pair of nodes $i$ and $j$ as a function of the product of their degrees $k_i$ and $k_j$. Same set of networks as in Fig.~\ref{fig:Davg_vs_k}.
For real-world networks, the result of \Pkk-theory (left column) is more accurate than the result of $p_k$-theory (right column).}
\label{Davg_vs_kk}
\end{figure}

As our last example, we show in Fig.~\ref{Davg_vs_kk} the expected distance between a random pair nodes as a function of the product of their degrees. We use the same set of networks as in Fig.~\ref{fig:Davg_vs_k}, and plot
\begin{align}
 \lb<D_{k_i k_j}\rb>=\frac{\sum_{k_i'\le k_j'}\delta_{k_i' k_j',k_i k_j} \frac12 \lb(\Dbar^{k_i',k_j'} + \Dbar^{k_j',k_i'}\rb) }{\sum_{k_i'\le k_j'} \delta_{k_i' k_j',k_i k_j}}, \label{e:Dkk}
\end{align}
where $\Dbar^{k,k'} = \sum_n n D_n^{k,k'}$, and $\delta_{i,j}$ is the Kronecker delta function. As in Fig.~\ref{fig:Davg_vs_k}, \Pkk-theory usually works better than $p_k$-theory for real-world networks. For synthetic networks the theoretical results are again in excellent agreement with the exact numerical calculations.

Finally, we note that theoretical predictions for networks that contain degree-1 nodes can be further improved using the procedure described in Appendix~\ref{App:rectified}.

\section{Conclusions} \label{s:conclusions}
We have presented a simple yet very powerful analytical method for calculating shortest path lengths in networks. Our approach is directly applicable to real-world networks with known degree distribution $p_k$ and/or joint degree-degree distribution \Pkk. It is simpler and yields more accurate predictions for synthetic and real-world networks than some previous analytical methods. Our approach can be extended to modular networks~\cite{Gleeson08a,Melnik14}, to networks with non-zero clustering~\cite{Gleeson09a,Gleeson09b,Hackett11,Miller09b,Newman09}, and also to directed networks~\cite{Gleeson08b}. We hope that these results will be useful for investigating the interdependence of network characteristics and will help advance the understanding of network structure and dynamics.

\section*{ACKNOWLEDGEMENTS}
S.M. acknowledges the INSPIRE fellowship funded by the Irish Research Council co-funded by Marie Curie Actions under FP7. J.P.G. and S.M. acknowledge funding provided by Science Foundation Ireland under programmes 11/PI/1026 and 12/IA/1683. We thank Ali Faqeeh, Dmitri Krioukov, Sergey Dorogovtsev and Alexander Goltsev for useful discussions. We thank Mason Porter, Adam D'Angelo and Facebook for providing the Facebook data used in this study. We also thank Jukka-Pekka Onnela, Alex Arenas, Mark Newman, and Cx-Nets collaboratory for sharing other data sets used in this paper. This work was conceived in part at the Complex Systems Summer School (CSSS) in Santa Fe Institute, NM, USA.

\appendix

\section{Analytical results for $z$-regular random graphs} \label{App:RRG}
In this Appendix, we obtain analytical formula \eqref{LargeNRRG} for the distribution of distances in $z$-regular graphs by solving Eqs.~\eqref{RRG_rhon}--\eqref{RRG_qn}. For comparison, we obtain similar analytical formulas based on the results of other authors~\cite{Fronczak05,Dorogovtsev03b}.

\subparagraph{Our formula:}
Substituting $q_n = 1- y_n$ into Eq.~\eqref{RRG_qn}, we have 
\begin{align}
y_n = (1-\rho_0) (y_{n-1})^{z-1}
\end{align}
with $y_0 = 1-\rho_0$. This can be solved exactly (let $u_n = \ln y_n$ and $u_n$ solves the linear difference equation $u_n = \ln(1-\rho_0) + (z-1)u_{n-1}$, with solution of form $u_n = A \lambda ^n + B$) to yield
\begin{align}
 y_n = \exp\lb[\frac{(z-1)^{n+1}-1}{z-2} \ln (1-\rho_0)\rb]. \label{yn}
\end{align}
Using~\eqref{Dn} and~\eqref{RRG_rhon}, we can write $D_n$ in terms of $y_n$ as follows:
\begin{align}
\nn D_n &= \rho_n - \rho_{n-1} \\
&= (1-\rho_0) \lb[ y_{n-2}^z - y_{n-1}^z \rb],
\end{align}
and inserting the solution~\eqref{yn} for $y_n$ gives
\begin{align}
\nn D_n =& \exp \lb[ \frac{z(z-1)^{n-1}-2}{z-2}\ln(1-\rho_0) \rb]-\\ 
&\exp \lb[ \frac{z(z-1)^n-2}{z-2}\ln(1-\rho_0) \rb]. 
\end{align}
Next, using the fact that $\ln(1-\rho_0)=\ln\lb(1-\frac{1}{N}\rb)\approx -\frac{1}{N}$ for large $N$ (with error of order $1/N^2$), we obtain Eq.~\eqref{LargeNRRG}:
\begin{align}
\nn D_n^{\rm RRG} = \exp \lb[ -\frac{z(z-1)^{n-1}-2}{(z-2)N} \rb] - \exp \lb[ -\frac{z(z-1)^n-2}{(z-2)N} \rb].
\end{align}

\subparagraph{Formula based on Fronczak et al.~\cite{Fronczak05}:}
Using the fact that all nodes have degree $z$ in Eq.~(10) of~\cite{Fronczak05} and substituting the result in Eqs.~(9) and (4) of~\cite{Fronczak05}, we obtain 
\begin{align}
 D^{\rm FR.}_n = \exp \lb[ \frac{-z(z-1)^{n-2}}{N} \rb] - \exp \lb[ \frac{-z(z-1)^{n-1}}{N} \rb].
 \label{FronczakRRG}
\end{align}

\subparagraph{Formula based on Dorogovtsev et al.~\cite{Dorogovtsev03b}:}
Here we follow the procedure described in Sec.~6 of~\cite{Dorogovtsev03b} to derive an explicit analytical formula for the \DD~for $z$-regular random graphs. Unfortunately it is not possible to follow this procedure for an arbitrary degree distribution, so we only use $z$-regular networks to compare the results of~\cite{Dorogovtsev03b} with our own. 

For $z$-regular networks with degree distribution $p_k=\delta_{k,z}$, we have $z_0=z$ and $z_1=z-1$.

Step 1: $\phi(x)=x^z$, $\phi_1(x)=\frac{\phi'(x)}{\phi'(1)}=x^{z-1}$.

Step 2: $f(y)$ solves $f(z,y)=\phi_1[f(y)]$, hence $f(z,y)=[f(y)]^{z-1}$ with $f(0)=1, f'(0)=-1$. Solution: $f(y)=e^{-y}$.

Step 3: $g(y):=\phi(f(y))=e^{-zy}$.

Step 4: $p(x)$ is inverse Laplace transform of $g(y)$ hence $p(x)=\delta(x-z)$.

Next, substituting $g$ and $p$ into Eqn.~(39) of~\cite{Dorogovtsev03b} yields ${\cal Q}(l)=\int_0^\infty dx \, p(x)[1-g(z_1^lx)]=1-g(z z_1^l)=1-e^{-z^2 z_1^l}$. Then $Q_n = {\cal Q}(n-n_0)=1-e^{-z^2 z_1^{n-n_0}}$ is the probability that two randomly chosen nodes are separated by a distance $\le n$. From Eqn. (34) of~\cite{Dorogovtsev03b}, $n_0=\ln_{z-1}(z(z-2)N)$ hence $(z-1)^{-n_0}=\frac{1}{z(z-2)N}$. Thus
\begin{align}
\nn D_n^{\rm DOR.} &= Q_n-Q_{n-1} = e^{-z^2(z-1)^{n-1-n_0}}-e^{-z^2(z-1)^{n-n_0}}\\
&=\exp \lb[ -\frac{z(z-1)^{n-1}}{(z-2)N}\rb]- \exp \lb[ -\frac{z(z-1)^n}{(z-2)N}\rb].
\end{align}

\section{Analytical results for the average shortest path length} \label{App:Davg}
In this Appendix, we list some previously obtained analytical results for the average shortest path length in random networks with arbitrary degree distribution $p_k$ and $N$ nodes. In the equations below, $\lb< \bullet \rb>$ denotes averaging with respect to $p_k$.

\subparagraph{Average shortest path length by Fronczak et al.~\cite{Fronczak05}:}
Equation~(29) of~\cite{Fronczak05} gives
\begin{align}
 \Dbar^{\rm FR.} = \frac{-2 \lb< \ln k \rb> + \ln \lb( \lb< k^2 \rb> - \lb< k \rb> \rb) + \ln N -\gamma}{\ln \lb( \lb< k^2 \rb> / \lb< k \rb> -1 \rb) } + \frac12, \label{e:Dbar_FR}
\end{align}
where $\gamma\approx 0.5772$ is the Euler-Mascheroni constant. The result of~\eqref{e:Dbar_FR} for $z$-regular random graphs is indistinguishable from that obtained using~\eqref{FronczakRRG} in $\sum_n n D_n$.

\subparagraph{Average shortest path length by Newman et al.~\cite{Newman01a}:}
Equation~(53) of~\cite{Newman01a} gives
\begin{align}
 \Dbar^{\rm NMN.} = \frac{\ln((N-1)(z_2-z)+z^2) - \ln z^2}{\ln z_2/z}, \label{e:Dbar_NMN}
\end{align}
where $z$ is the mean degree, $z_2$ is the mean number of second neighbors. For random networks with degree distribution $p_k$, we have $z_2=\lb<k^2\rb>-\lb<k\rb>$, hence for $z$-regular random graphs $z_2 = z^2-z$, and for \ER~networks $z_2 = z^2$.

\section{Results of Katzav et al. for \ER~graphs} \label{App:Katzav}
In~\cite{Katzav15}, Katzav et al.~calculate the \DD~for an \ER~network with $N$ nodes where each pair of nodes is connected with probability $p$, i.e., an \ER~network with mean degree $z=Np$. The probability $F_n$ that the distance between a random pair of nodes in the same connected component is larger than $n$ can be calculated from Eqs.~(2) and (3) of~\cite{Katzav15}:
\begin{align}
F_{n} &= F_1 \prod_{m=2}^n P(N,m),
\end{align}
where $F_1=1-p$, and $P(N,m)$ can be found from the recurrence equation
\begin{align}
 P(N,m) &= \lb(1-p + p P(N-1,m-1)\rb)^{N-2},
\end{align}
starting with $P(N-m+1,1)=1-p$. Then, the \DD~is given by 
\begin{align}
D_n^{\rm KAT.} &= F_{n-1}-F_n. \label{Katzav}
\end{align}

\section{Intrinsic mismatch between single and multiple seed cascades} \label{s:error_pk}
Here we explain the main reasons for less than perfect (though still excellent) performance of $p_k$-theory for random networks with degree distribution $p_k$. 

Our method is based on the fact that an epidemic process started with a single seed as described in Sec.~\ref{s:analogy} can be used to calculate the exact \DD~between network nodes. Our analytical approach (Sec.~\ref{s:z_regular}-\ref{Pkk_theory}) to solving this epidemic process assumes, like many other theories, that the network is infinite. Therefore, the initial infection of a single randomly-chosen degree-$k'$ seed node in a finite network of $N$ nodes with degree distribution $p_k$ is represented in our approach as the infection of a small (finite) fraction of nodes $\rho_0^{k,k'}$ (given by Eq.~\eqref{rho0}) in an infinite network. This means that our $p_k$-theory of Sec.~\ref{pk_theory} actually describes an epidemic with a very large number $m$ of degree-$k'$ seeds in a very large network with $mN$ nodes. Therefore, after the first update of nodes' states, since the nodes are connected at random, $p_k$-theory predicts non-zero fractions of infected nodes of all degrees in the network. In reality, however, a single degree-$k'$ seed node cannot infect more than $k'$ of its neighbors, so there will be at most $k'$ degree classes with infected nodes. This difference between $p_k$-theory prediction and a single-seed epidemic is easily seen when $k'$ is small, and to some extent it affects the results for all time steps.

\begin{figure}[htb]
\centering
\includegraphics[width=0.97\columnwidth]{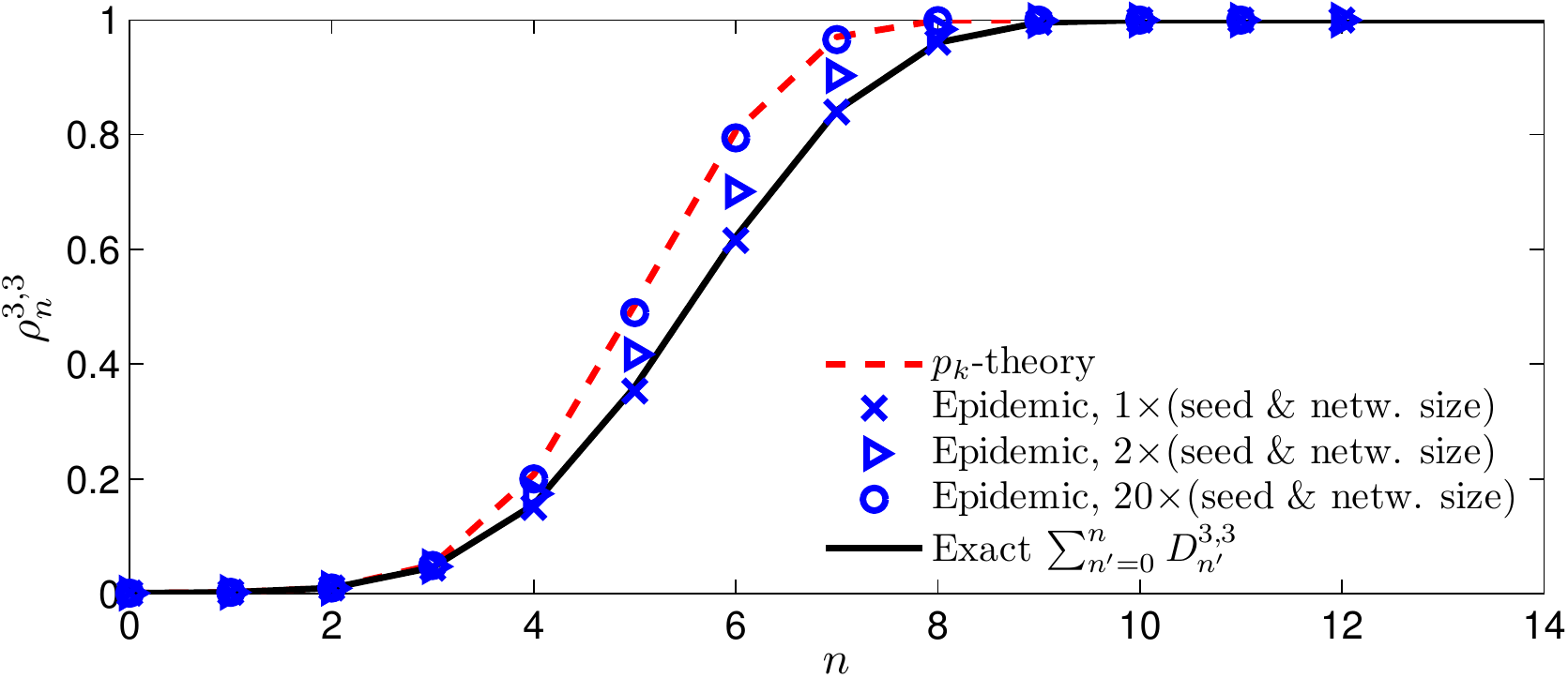}
\caption{(Color online) Evolution of the expected infected fraction of degree-$3$ nodes for an epidemic started by degree-$3$ node(s) on a $(3,50)$-regular random network with nodes in proportion $p_3/p_{50}=200/1$. 
We specifically chose this network to emphasize the difference between $p_k$-theory results (red dashed line) and the exact evolution (black solid line). The exact evolution is the cumulative sum $\sum_{n'=0}^n D_n^{3,3}$ of the actual \DD~on the network. The expected value of the epidemic started with a single seed (blue crosses) matches the exact results (black solid line). As we increase the number of seed nodes in the epidemic (together with the network size, so that the seed fraction is unchanged), the epidemic results converge to $p_k$-theory results. Numerical results for epidemics are averaged over 1000 random sets of seeds.}
\label{f:3_50_RRG_Nseeds}
\end{figure}

To illustrate this point, in Fig.~\ref{f:3_50_RRG_Nseeds} we consider a $(3,50)$-regular random network~\cite{Melnik13,Melnik14}, which consists of nodes of degrees 3 and 50 in proportion 200:1. We compare the exact results with the prediction of $p_k$-theory and with the simulations of epidemic processes with single or multiple seeds. We specifically chose such a network to observe a clear difference between the exact results and $p_k$-theory, because in most other situations the difference is much less prominent.

We take $N=2000$ nodes in Fig.~\ref{f:3_50_RRG_Nseeds} and plot the time evolution of the expected fraction of degree-3 nodes infected by a degree-3 seed. The exact time evolution (black solid line) can be obtained by taking the cumulative sum $\rho_n^{3,3}=\sum_{n'=0}^n D_{n'}^{3,3}$ of the \DD~$D_n^{3,3}$ between degree-3 nodes.\footnote{$D_n^{3,3}$ can be calculated, for example, using the Dijkstra algorithm.} The expected value of the epidemic started with a single seed (blue crosses) matches the exact result, but not the result of $p_k$-theory (red dashed line). As we increase the number of seed nodes and $N$ by a factor of 2 and 20 (blue triangles and circles), the epidemic results converge to $p_k$-theory results. We increase $N$ by the same factor as the number of seed nodes to keep unchanged the fraction of seed nodes $\rho_0^{k,k'}$. The result of $p_k$ theory remains the same as it depends only on the fraction of seed nodes $\rho_0^{k,k'}$ and the network degree distribution $p_k$.

One way to improve the prediction of $p_k$-theory is to further partition nodes into ``types'' based on the degrees of their first, second, third, etc. neighbors. For example, in the $(3,50)$-regular random network, degree-$3$ nodes can be split into two node types: those who are neighbors of degree-$50$ nodes, and those who are not. In two time steps, the first type infects many more nodes than the second one, but such differences are not taken into account by $p_k$-theory because it describes nodes based only on their degrees. Considering node types within each degree class will improve the theoretical prediction, but will necessarily lead to more complicated equations. We note that $p_k$-theory is extremely accurate for $z$-regular random graphs because there is only one node type and it is fully described by node degree.

\section{The difference between theoretical $\rho_n^{k,k'}$ and $\rho_n^{k',k}$} \label{s:asymmetry}
Here we explain the ``asymmetry'' in our theoretical results for $D_n^{k,k'}$, i.e., the fact that $D_n^{k,k'}$ and $D_n^{k',k}$ are equal only approximately because $\rho_n^{k,k'}\ne \rho_n^{k',k}$.

Consider a $(z_1,z_2)$-regular random network~\cite{Melnik13,Melnik14}, which consists of $N$ nodes of degrees $z_1$ and $z_2$; the fractions of nodes of each degree are $p_{z_1}$ and $p_{z_2}$ respectively. There are two ways to calculate the \DD~between $z_1$ and $z_2$ nodes. First, we can consider $\rho_n^{z_2,z_1}$, which is the time evolution of the expected fraction of infected degree-$z_2$ nodes in an epidemic started by a degree-$z_1$ seed; alternatively, we can start with a degree-$z_2$ seed and look at $\rho_n^{z_1,z_2}$. Numerical simulations of these single-seed epidemics on a given network will give us the same value $\rho_n^{z_2,z_1}=\rho_n^{z_1,z_2}$ for both cases, which yields the exact \DD~between $z_1$ and $z_2$ nodes. However, Eqs.~\eqref{pk_rhon}--\eqref{rho0} provide slightly different values for $\rho_n^{z_1,z_2}$ and $\rho_n^{z_2,z_1}$ because they operate with fractions of nodes in infinite networks, and therefore have no concept of a single seed node. 

For example, $\rho_1^{z_1,z_2}$ given by Eqs.~\eqref{pk_rhon}--\eqref{rho0} is the fraction of degree-$z_1$ nodes who are neighbors of the fraction $\rho_0^{z_2,z_2}$ of degree-$z_2$ nodes. Here, $\rho_0^{z_2,z_2}\equiv 1/(N p_{z_2})$ is the fraction of nodes (in an infinite network) that a single degree-$z_2$ node occupies in the network of interest. 

\begin{figure}[htb]
\centering
\includegraphics[width=0.97\columnwidth]{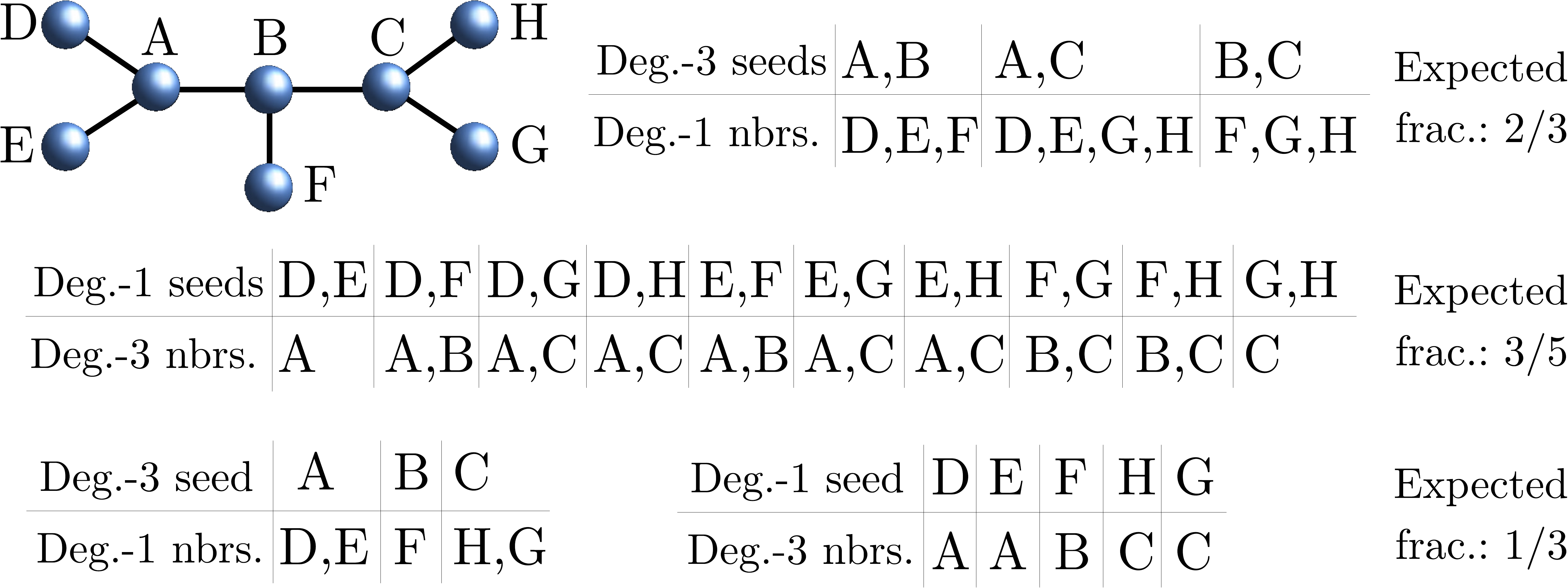}
\caption{(Color online) A network consisting of nodes of degree 1 and 3. Tables show the seed sets and the corresponding sets of neighbors of the other degree. For epidemics started by a pair of degree-3 seed nodes, the expected fraction of degree-1 nodes who are neighbors of the seeds is $(3/5+4/5+3/5)/3 = 2/3$. For epidemics started by a pair of degree-1 seeds, the expected fraction of degree-3 nodes who are neighbors of the seeds is $3/5$. However, for epidemics started by either a single degree-3 or a single degree-1 node, the fraction of neighbors of the other degree is 1/3 in both cases. This demonstrates that for epidemics started by more than one initially infected node, we cannot guarantee that $\rho_1^{z_2,z_1}=\rho_1^{z_1,z_2}$ and explains why Eqs.~\eqref{pk_rhon}--\eqref{rho0} generally produce slightly different values for $\rho_n^{k,k'}$ and $\rho_n^{k',k}$. }
\label{f:z1_z2_RRG_epidemic}
\end{figure}

To understand why dealing with fractions of nodes leads to inexact results, let us think of a direct numerical simulation that calculates $\rho_1^{z_1,z_2}$ (or $\rho_1^{z_2,z_1}$) for an epidemic described by Eqs.~\eqref{pk_rhon}--\eqref{rho0}. Such simulation should run on a very large network starting with an infected fraction $\rho_0^{z_2,z_2}\equiv 1/(N p_{z_2})$ of degree-$z_2$ nodes (which would involve many degree-$z_2$ nodes) to calculate $\rho_1^{z_1,z_2}$, the expected fraction of degree-$z_1$ who are neighbors of the infected degree-$z_2$ nodes. Similarly, $\rho_1^{z_2,z_1}$ can be obtained by initially infecting a fraction $\rho_0^{z_1,z_1}\equiv 1/(N p_{z_1})$ of degree-$z_1$ nodes (which would again be many nodes), and then calculating the expected fraction of degree-$z_2$ nodes who are neighbors of the infected nodes. The important point is that for epidemics starting with more than one initially infected node, we cannot guarantee that $\rho_1^{z_2,z_1}=\rho_1^{z_1,z_2}$. Consider, for example, the network in Fig.~\ref{f:z1_z2_RRG_epidemic}. Let us initially infect a pair of degree-$3$ (or a pair of degree-$1$) nodes and calculate the expected fraction nodes of the other degree who are neighbors of the infected pair. For a pair of degree-$3$ seeds we get 2/3, while for a pair of degree-$1$ seeds we get 3/5. Notice that both values are 1/3 if the seed is a single degree-$3$ or degree-$1$ node.

\section{Improving predictions for real-world networks} \label{App:rectified}
The following procedure further improves the prediction of $p_k$-theory for real-world networks (RWNs) that contain degree-1 nodes, though we found the improvement is small on the RWNs that we considered. If the degree distribution $p_k^{\rm RWN}$ of a RWN has degree-1 nodes, a random network (constructed using the configuration model) with $p_k^{\rm RWN}$ will have GCC size $<1$. In this case, the degree distribution of the GCC will be slightly different from $p_k^{\rm RWN}$ because nodes with low degree are less likely to be in the GCC comparing to high-degree nodes. Since our $p_k$-theory predicts the distribution of distances in GCC (which we then translate back to the real-world network), we want GCC to have the degree distribution $p_k^{\rm RWN}$. To achieve this, one should use a \emph{rectified} degree distribution $p_k^{\rm rec}$ in Eqs.~\eqref{pk_rhon}--\eqref{rho0} instead of the network's original degree distribution $p_k^{\rm RWN}$. A rectified degree distribution $p_k^{\rm rec}$ is the degree distribution that yields a random network whose GCC has the degree distribution $p_k^{\rm RWN}$. If there are no degree-1 nodes in the real-world network, then $p_k^{\rm rec}=p_k^{\rm RWN}$. One can find $p_k^{\rm rec}$ from $p_k^{\rm RWN}$ from the following equality
\begin{align}
 p_k^{\rm RWN} = F\lb(p_k^{\rm rec}\rb) \equiv \frac{p_k^{\rm rec} \rho_\infty^k}{\sum_k p_k^{\rm rec} \rho_\infty^k},
\end{align}
where $\rho_\infty^k$ are obtained by iterating to steady state Eqs.~\eqref{pk_rhon}--\eqref{pk_qbar} with $p_k^{\rm rec}$ and starting with infinitesimal $\rho_0^k$ and $q_0^k$ (note the dependence on $k'$ is absent here).

\begin{figure}[htb]
\centering
\includegraphics[width=0.97\columnwidth]{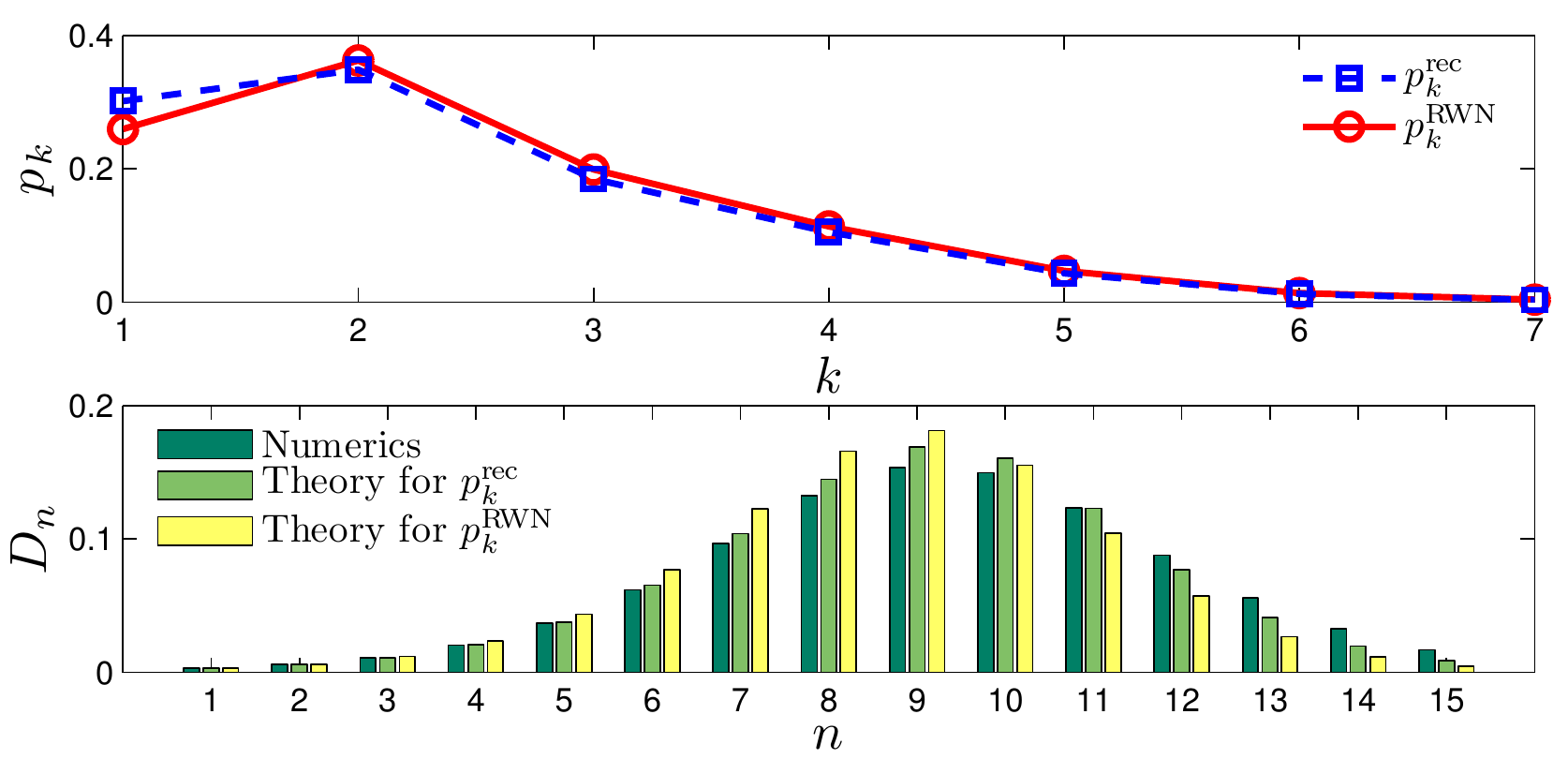}
\caption{(Color online) (Top) Comparison of the original $p_k^{\rm RWN}$ and rectified $p_k^{\rm rec}$ degree distributions. (Bottom) Distribution of shortest path lengths given by $p_k$-theory using 
$p_k^{\rm rec}$ matches better to the direct numerical calculation than that obtained using $p_k^{\rm RWN}$.} 
\label{f:pk_rectified}
\end{figure}

In Fig.~\ref{f:pk_rectified}, we demonstrate the advantage of using $p_k^{\rm rec}$ instead of $p_k^{\rm RWN}$. We generate an \ER~network consisting of $N=1000$ nodes with mean degree $z=2$, and treat its GCC as if it were a RWN. 

\bibliography{networks}
\end{document}